\documentclass[conference]{IEEEtran}

\usepackage[utf8]{inputenc}
\usepackage[T1]{fontenc}
\usepackage{microtype}
\usepackage[nolist]{acronym}

\usepackage{cite}
\usepackage{amsmath,amssymb,amsfonts}
\usepackage{algorithmic}
\usepackage{graphicx}
\usepackage{textcomp}
\usepackage{xcolor}
\usepackage{booktabs}
\usepackage{paralist}
\usepackage{comment}
\usepackage{soul}
\usepackage{tabularx}
\usepackage{amsthm}
\usepackage[hidelinks]{hyperref}
\usepackage{lscape} 
\theoremstyle{definition}
\newtheorem{definition}{Definition}[section]
\usepackage{varwidth}
\usepackage{tasks}

\def\baselinestretch{0.98}

\usepackage{enumitem}
\setitemize{noitemsep,topsep=0pt,parsep=0pt,partopsep=0pt, leftmargin=*}

\usepackage{csquotes}

\def\BibTeX{{\rm B\kern-.05em{\sc i\kern-.025em b}\kern-.08em
    T\kern-.1667em\lower.7ex\hbox{E}\kern-.125emX}}
\usepackage{xspace}

\newcommand{\ens}{\emph{Tax-DS}\xspace}
\newcommand{\pbr}{\emph{CrossVal-DS}\xspace}
\newcommand{\scrape}{\emph{General-DS}\xspace}

\DeclareUrlCommand\doi{\def\UrlLeft##1\UrlRight{DOI\nobreakspace\href{http://dx.doi.org/##1}{##1}}\urlstyle{rm}}

 \usepackage{amssymb}

\definecolor{aqua}{rgb}{0.0, 1.0, 1.0}

\begin{acronym}
    \acro{SE}{Software Engineering}
    \acro{RE}{Requirements Engineering}
    \acro{ML}{Machine Learning}
    \acro{TL}{Transfer Learning}
    \acro{BERT}{Bidirectional Encoder Representations from Transformers}
    \acro{NLP}{Natural Language Processing}
    \acro{DS}{Dataset}
    \acro{RQ}{Research Question}
    \acro{NB}{Naive Bayes}
    \acro{DT}{Decision Tree}
    \acro{TF-IDF}{Term Frequency–Inverse Document Frequency}
    \acro{BoW}{Bag of Words}
    \acro{KNN}{K-Nearest Neighbor}
    \acro{SVM}{Support Vector Machine}
    \acro{RF}{Random Forest}
    \acro{LR}{Logistic Regression}
    \acro{AB}{AdaBoost}
    \acro{DL}{Deep Learning}
    \acro{PAD}{padding tokens}
    \acro{CLS}{classification token}
    \acro{FN}{False Negative}
    \acro{FP}{False Positive}
\end{acronym}

\usepackage{amssymb}
\usepackage{environ}

\NewEnviron{myequation}{%
\begin{equation}
\scalebox{0.9}{$\BODY$}
\end{equation}
}

\usepackage{multirow}
\usepackage[table]{colortbl}

\begin{document}

\title{Explanation Needs in App Reviews: Taxonomy and Automated Detection}

\makeatletter
\newcommand{\linebreakand}{%
  \end{@IEEEauthorhalign}
  \hfill\mbox{}\par
  \mbox{}\hfill\begin{@IEEEauthorhalign}
}
\makeatother


\author{\IEEEauthorblockN{Max Unterbusch}
\IEEEauthorblockA{
\textit{University of Cologne} \\
munterbusch@smail.uni-koeln.de}
\and
\IEEEauthorblockN{Mersedeh Sadeghi}
\IEEEauthorblockA{
\textit{University of Cologne} \\
sadeghi@cs.uni-koeln.de}
\and
\IEEEauthorblockN{Jannik Fischbach }
\IEEEauthorblockA{
\textit{Netlight Consulting GmbH | fortiss GmbH} \\
jannik.fischbach@netlight.com}
\linebreakand
\IEEEauthorblockN{Martin Obaidi}
\IEEEauthorblockA{
\textit{Leibniz University Hannover, Software Engineering Group}\\
martin.obaidi@inf.uni-hannover.de}
\and
\IEEEauthorblockN{Andreas Vogelsang}
\IEEEauthorblockA{
\textit{University of Cologne} \\
vogelsang@cs.uni-koeln.de}
}

\maketitle

\begin{abstract}
Explainability, i.e. the ability of a system to explain its behavior to users, has become an important quality of software-intensive systems. Recent work has focused on methods for generating explanations for various algorithmic paradigms (e.g., machine learning, self-adaptive systems). 
There is relatively little work on what situations and types of behavior should be explained. There is also a lack of support for eliciting explainability requirements.
In this work, we explore the need for explanation expressed by users in app reviews. We manually coded a set of 1,730 app reviews from 8 apps and derived a taxonomy of \emph{Explanation Needs}.
We also explore several approaches to automatically identify \emph{Explanation Needs} in app reviews. Our best classifier identifies \emph{Explanation Needs} in 486 unseen reviews of 4 different apps with a weighted F-score of 86\%.
Our work contributes to a better understanding of users' \emph{Explanation Needs}. Automated tools can help engineers focus on these needs and ultimately elicit valid \emph{Explanation Needs}.
\end{abstract}


\section{Introduction}
Software systems are becoming more intelligent and ubiquitous than ever before, increasing the criticality of their impact on humans. Driven by modern artificial intelligence, it is becoming increasingly difficult for an external user, but also for the developers of these systems, to understand their inner workings and thus their decisions and actions. The ability to provide explanations---a natural ability of humans---is therefore considered an important capability of software systems. As such, explainability is now accepted as a critical quality attribute~\cite{chazette_explainability_2020} and represents an emerging topic in the field of \ac{RE}~\cite{brunotte_quo_2022}.

Researchers have explored the foundations of explainability from different angles. There are several approaches to generating explanations for different algorithmic paradigms.
However, there has been relatively little focus in the literature on what users actually need explanations for~\cite{sadeghi_cases_2021}. 
This lack of knowledge limits our ability to effectively elicit explainability requirements and apply existing explanation generation methods. 
Thus, the first problem we address in this paper is as follows: \textit{We lack knowledge about what users need explanations for}. 

App reviews have been overlooked as a potential source of \emph{Explanation Needs}. Pagano and Maalej~\cite{pagano_user_2013} found that app reviews contain valuable \ac{RE}-related information because they represent rich and readily available textual data that provides insights into thousands of user experiences. 
Unlike interview or survey data, app reviews are collected ``in the field'' under natural circumstances. Users are motivated enough to publish their opinions about an app; they are not forced or paid to do so. This underlines the importance that users place on their concerns. In addition, users are not asked about any specific aspect. The review messages are open to any feedback the users want to give to the app vendors or developers.

We set out to understand users' need for explanation, which we refer to as \emph{Explanation Need}. Our focus is to characterize the occurrence of \emph{Explanation Needs} in app reviews and to investigate the types of \emph{Explanation Needs} that users express. We conducted a qualitative analysis of 1,730 English app reviews of 8 different apps. As a result, we propose a taxonomy of \emph{Explanation Needs} in app reviews to help developers and researchers distinguish between different types. One of the key benefits of the taxonomy is that it enables researchers and engineers to extract explainability requirements in a systematic and rigorous manner. By categorizing users' \emph{Explanation Needs} from their perspective into distinct categories, the taxonomy highlights areas where a system may lack transparency or fail to meet users' expectations. This, in turn, provides valuable insight into the types of explanations that are most needed.

Our qualitative analysis shows that \emph{Explanation Needs} in app reviews are valuable and contain rich information, but are relatively sparse. \emph{Explanation Needs} have only appeared in about 5\% of the app reviews studied. However, manually analyzing app reviews can be challenging due to the sheer volume of reviews and the varying levels of detail and insight they provide.
Tool support to filter the reviews for relevant content would be valuable to allow development and stakeholders to efficiently exploit this source of information~\cite{guzman_how_2014}. We identify this as the second problem addressed in this paper: \textit{We lack tool support to automatically identify \emph{Explanation Needs} in app reviews}. To support the use of app reviews, we investigated several classifiers (rule-based, traditional \ac{ML}, and transformer approaches) to automatically detect \emph{Explanation Needs} in app reviews. We evaluated and compared the classifiers in a 10-fold cross-validation on an extended set of 5,078 manually labeled app reviews. In addition, we evaluated our baseline rule-based approach and our best-performing classifier on an additional set of 486 unseen and unmodified reviews of 4 new apps to test how well the approaches generalize and perform in a realistic setting.
Our best-performing classifier---a fine-tuned \ac{BERT} model---achieved a weighted F-score of 93\% in a 10-fold cross-validation and a weighted F-score of 86\% when evaluated on unseen data. We make the following contributions:
\begin{itemize}[noitemsep,topsep=0pt,parsep=0pt,partopsep=0pt]
\item We provide a taxonomy of \emph{Explanation Needs} derived from a large set of app reviews. 
\item We provide a performance analysis of several classifier approaches to detect \emph{Explanation Needs} automatically in app reviews. 
\item We publish a set of 5,564 app reviews that we manually labeled according to our proposed taxonomy.  
\item To strengthen transparency and facilitate replication, we make our code, dataset, and trained models publicly available.\footnote{\doi{10.5281/zenodo.7740411  }.}
\end{itemize}

\section{Terminology and Related Work}


\subsection{Explainability and User Needs in Explanations}\label{subsec:expl-user-needs}

Explainability has gained significant attention from various research fields, including Human-Computer Interaction, Cyber-Physical Systems, and Psychology~\cite{brunotte_quo_2022}. Since 2019, when it was proposed as a non-functional requirement ~\cite{kohl_explainability_2019}, it has become a trending topic within the \ac{SE} and \ac{RE} communities~\cite{brunotte_quo_2022}. Research has shown that explainability can enhance trustworthiness, transparency, accountability, fairness, ethics, and other quality aspects by overcoming the black box nature of software systems ~\cite{chazette_exploring_2021, kastner_relation_2021, leite_software_2010}. Chazette et al. developed a concise definition of explainability that meets the requirements of \ac{SE} and \ac{RE} communities~\cite{chazette_exploring_2021}: \enquote{A system S is explainable with respect to an aspect X of S relative to an addressee A in context C if and only if there is an entity E (the explainer) who, by giving a corpus of information I (the explanation of X), enables A to understand X of S in C.} The explainer entity does not have to be the system itself. Achieving explainability depends on specific variables: the system's aspect, the addressee, and the context. Accordingly, Kohl~\cite{kohl_explainability_2019} and Chazette~\cite{chazette_explainability_2020} emphasize the significance of identifying users' specific needs for explanations and providing customized explanations correspondingly. Indeed, in cases where users do not require explanations, ensuring explainability may not be necessary~\cite{chazette_explainability_2020}.

Studying app reviews for explanation need identification is a relatively under-researched area. Consequently, a taxonomy of \emph{Explanation Needs} can aid in advancing knowledge and eliciting requirements for developing explainable systems. Constructing taxonomies provides numerous benefits, including supporting the communication of complex concepts, revealing relationships between entities, and uncovering knowledge gaps. In a similar approach for a different domain, Sadeghi et al.~\cite{ sadeghi_cases_2021} developed a taxonomy of reasons for \emph{Explanation Needs}. They primarily distinguish between four categories of situations requiring explanations: \textit{Training}, \textit{Interaction}, \textit{Debugging}, and \textit{Validation}, yet the authors focused on \textit{Interaction}. For \textit{Interaction}, the taxonomy further breaks hierarchically down into disobedience, failure, and context-aware behavior. That work considered the system, the user, and the environment in their taxonomy; in contrast, our focus will be on the user only.

\subsection{App Store Mining and Classifying App Reviews}\label{subsec:app-store-mining}

Pagano et al.~\cite{pagano_user_2013} conducted a
comprehensive analysis of app stores to determine their usefulness for
requirements engineering. They collected over a million app reviews and found that feedback messages can facilitate communication between users and developers. However, they discovered that a significant amount of the feedback collected was of poor quality and lacked informative value. They argue that although app stores can facilitate user-centered \ac{RE} through the use of user feedback, it is essential to employ appropriate tools and techniques to filter and pre-process relevant contributions. In response to the need for tool support in app store mining, the \ac{RE} community developed various solutions to extract valuable insights from app store reviews. Guzman and Maalej~\cite{guzman_how_2014} proposed a method to filter features mentioned by users and extract corresponding sentiments, allowing for a detailed analysis of user experience with individual app features. Chen et al. presented a tool that filters app reviews, groups and ranks them, and provides visualizations of the insights~\cite{chen_ar-miner_2014}.

Particularly relevant to this paper are contributions that classify app reviews according to predefined labels, such as problem reports, inquiries, and user experience, or non-functional requirements such as reliability, usability, and portability. To achieve this classification, researchers typically use traditional \ac{ML} and \ac{DL} methods for classifying app reviews into various categories ~\cite{ guzman_ensemble_2015, maalej_automatic_2016, maalej_bug_2015}. Active Learning strategies have also been experimented with, which can help reduce human labor and improve classification accuracy in certain scenarios ~\cite{dhinakaran_app_2018}.
Recently, \ac{BERT} achieved state-of-the-art performance classifying English app reviews into feature requests, problem reports, and irrelevant~\cite{henao_transfer_2021}. In this paper, we compare a simple rule-based approach as a baseline, different \ac{ML}-based approaches, and a \ac{DL}-based approach using the \ac{BERT}-Base model~\cite{devlin_bert_2019} for detecting \emph{Explanation Needs} in reviews automatically. 

\section{Characterization of Explanation Needs}\label{sec:taxonomy}

We define an \emph{Explanation Need} as a knowledge gap that a user intends to close and present our findings on such needs in app reviews in this section. To consider a review as an \emph{Explanation Needs}, the user must explicitly raise a question or express a need for an explanation. Rhetorical questions ([sic] ``What the hell?'') do not qualify as \emph{Explanation Needs} as they are not intended to elicit an answer. Direct requests ([sic] ``Please could you please check it?'') are also excluded since they do not indicate a specific gap in knowledge. 
It is important to note that we distinguish between \emph{Explanation Needs} and \emph{Explainability Need}, a non-functional requirement identified for software systems. On the other hand, \emph{Explanation Needs} are needs perceived by users. Following the formatting of Chazette et al.'s definition of explainability~\cite{chazette_exploring_2021}, we formally define \emph{Explanation Needs} as:

\begin{definition}[Explanation Needs]\label{def:exp_need} 
    An addressee A has incomplete knowledge about an aspect X of system S in context C and requests a corpus of information I provided by an entity E that allows A to understand X of S in C.
\end{definition} 


\subsection{Study Design}\label{subsec:tax-studydesing}
In the endeavor to identify users' \emph{Explanation Needs}, this research aims to explore the potential of app reviews as a source of information. By analyzing the rich textual data of reviews, we seek to uncover the types of explanations that users are looking for. To guide our investigation, we formulated the following research questions (RQ):


\noindent\textit{\textbf{RQ1:}} What types of \emph{Explanation Needs} have been expressed in app reviews?

\noindent\textit{\textbf{RQ2:}} How prevalent are \emph{Explanation Needs} and their types in app reviews? 

\begin{figure}
    \centering
    \includegraphics[width=\linewidth]{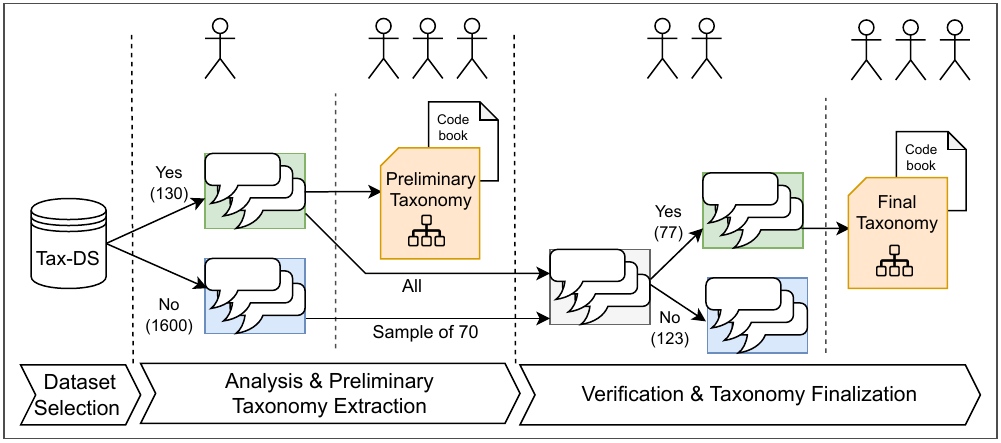}
    \caption{Taxonomy Creation Procedure.}
    \label{fig:overview_qualitative_analysis}
            \vspace{-.4cm}
\end{figure}

Answering \textit{RQ1 }is crucial for identifying common issues faced by users and prioritizing areas for improvement in app development. It aims to identify and understand users' \emph{Explanation Needs} in app reviews, guiding the development of more transparent and user-friendly software systems. To answer our research question, we undertake a qualitative analysis to develop a taxonomy for \emph{Explanation Needs} in app reviews. The provision of conception classification and taxonomy is generally valuable since it provides a standardized framework and facilitates a common ground to communicate and research in emerging fields of knowledge~\cite{vegas_maturing_2009}. As depicted in Figure~\ref{fig:overview_qualitative_analysis}, the qualitative analysis toward addressing \textit{RQ1} involved three phases: (1) Dataset Selection, (2) Analysis and Preliminary Taxonomy Extraction (3) Verification and Taxonomy Finalization.

\noindent\textbf{Phase 1.}
In the first phase, we selected the datasets for our analysis.
The original dataset used in our study was assembled by Brunotte~\cite{brunotte2022app}. Although a more recent version of the dataset exists with a larger number of reviews, we focused our analysis on a subset of 1,730 reviews provided to us directly by the authors. It allowed us to conduct our analysis more targeted and manageable. In the remainder of this paper, we refer to this dataset as \ens. \ens comprise app reviews from eight distinct apps available on the Apple App Store and Google Play. The domains represented in \ens span several categories, including health and wellness, finance, technology, and lifestyle, making it well-suited for exploring the nature of user feedback and \emph{Explanation Needs} in mobile app reviews. Table~\ref{tab:overviewApps} provides an overview of this dataset.

\noindent\textbf{Phase 2.}
Using the dataset as our basis, we extracted the preliminary taxonomy of \emph{Explanation Needs}. A single coder initially analyzed all 1,730 app reviews based on the definition of \emph{Explanation Needs} outlined in~\ref{def:exp_need}. The coder then filtered out 1,600 reviews that did not express any \emph{Explanation Need}, and the remaining 130 cases were labeled as \emph{Explanation Need} on a tentative basis. While there was a possibility that some of these cases could be excluded by the other coders in subsequent phases, these 130 cases still provided a foundation for further analysis in terms of categorization and taxonomy extraction. Following the template by Saldaña~\cite{saldana_coding_2013}, the coder also developed a codebook to maintain, organize, and share the codes with the other authors.
The initial coding resulted in an early version of the taxonomy, which was subject to further refinement through extensive discussions and revisions by the authors involved in the study. Hence, as this phase's output, a preliminary taxonomy was generated, which classified different types of \emph{Explanation Needs} and established boundaries between them. Nevertheless, at this point, the codebook yet had rather generic and fuzzy definitions of the categories or loose criteria for differentiating them. Therefore, we proceed to the next phase to further verify the applicability of the taxonomy and codebook.\looseness=-1

\noindent\textbf{Phase 3.}
In the final phase, we aimed to verify and refine the preliminary taxonomy by involving two other coders. We sampled 130 app reviews tentatively identified as \emph{Explanation Needs} by the first coder, plus a random selection of 70 reviews that were not labeled as such. The resulting dataset was shuffled and divided equally between the coders, with each responsible for categorizing their respective half as \emph{Explanation Need} or not. For the reviews categorized as \emph{Explanation Need}, the coders then had to check if they could be classified under one of the leaf nodes of the preliminary taxonomy. The goal was to ensure the preliminary taxonomy and codebook's completeness and accuracy and identify any deficiencies.
The coders then engaged in several rounds of discussions and classification. During the first iteration, the coders compared the labels assigned by the initial coder to the new labels the additional coders gave. From the 130 cases identified by the initial coder as \emph{Explanation Need}, 48 cases were excluded by either of the new coders. So we were left with 82 app reviews that the new coders also tentatively labeled as \emph{Explanation Need}, with each case being assigned a specific type of explanation. During the second iteration, all the coders went through these 82 reviews to further discuss and evaluate each case.
Moreover, at this point, coders attempted to prune and/or extend the taxonomy categorization to produce the final taxonomy and to consolidate their descriptions and boundaries recorded in the codebook. Throughout the last iteration, 5 additional app reviews that did not meet the requirements and specifications of the final taxonomy were excluded, resulting in a total of 77 cases labeled as \emph{Explanation Need}.

\begin{figure}
    \centering
    \includegraphics[width=0.8\linewidth]{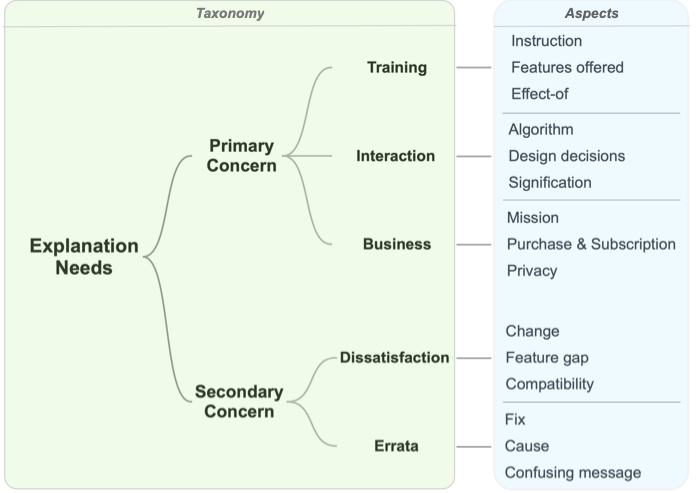}
            \vspace{-.1cm}
    \caption{Taxonomy of Explanation Needs in App Reviews}
    \label{fig:Taxonomy}
        \vspace{-.4cm}
\end{figure}

\subsection{Results: A Taxonomy of Explanation Needs}\label{subsec:tax}
As shown in Figure~\ref{fig:Taxonomy}, the taxonomy has a hierarchical structure and consists of two levels. We refer to the lowest level elements, namely \textsf{\textit{Training}}, \textsf{\textit{Interaction}}, \textsf{\textit{Business}}, \textsf{\textit{Dissatisfaction}}, and \textsf{\textit{Errata}}, as \textit{categories} of \emph{Explanation Needs}. To make the categories more tangible, we included a non-exhaustive list of \underline{\textit{aspects}} for each category. These aspects are more concrete groupings of related and typical \emph{Explanation Needs} that we could observe in the data. However, they are not part of the taxonomy in a narrow sense.

Given the \emph{Explanation Need}~\ref{def:exp_need}, a key distinction we make in the first level of our taxonomy is whether such a need for some explanation is an issue's primary or secondary concern. More precisely, if the user perceives their lack of knowledge as the only issue, then the \emph{Explanation Need} becomes a \textit{Primary Concern}, whereas if they see other substantial problems aside from their knowledge gap, it becomes a \textit{Secondary Concern}. In the latter case, an underlying problem exists, typically a deficiency, which substitutes the \emph{Explanation Need} as the primary concern. Therefore, offering an explanation may increase the overall understanding of the situation, but an explanation alone cannot solve the underlying problem. 

As depicted in Figure~\ref{fig:Taxonomy}, the \emph{Explanation Need}s belonging to the \textsf{\textit{Training}}, \textsf{\textit{Interaction}}, and \textsf{\textit{Business}} categories represent a primary concern. In general, \textsf{\textit{Training}} is when users are unfamiliar with the system or particular features, either because they are new to it or the system's features have been changed. We found the following aspects to characterize \textsf{\textit{Training}} best:

\begin{itemize}[noitemsep,topsep=0pt,parsep=0pt,partopsep=0pt, leftmargin=*]
    \item \underline{\textit{Instruction}}. Users seek instructions for achieving specific goals, such as how to use a system, feature, or settings option. This aspect requires that the users clearly intend what they aim to do. \textit{Instruction} aspect excludes reviews if there is an identifiable deficiency, such as an error or failure (see aspect \textit{Fix}). Example [sic]: ``How do you edit from this app???''.
      
    \item \textit{\underline{Features Offered.}} Users seek information about specific or general systems' features or functionality. Therefore, users are unaware of what the system can exactly do. Example [sic]: ``... is there anyway to sort this out ...?''.
    
    \item \textit{\underline{Effect-Of.}} Users want to obtain information on the potential outcomes of specific actions. The users know how to perform such an action but are not sure what the impact will be. Example [sic]: ``If I invest in dividend paying stocks, will the dividends be added to my portfolio?''. 
\end{itemize}

The next category is  \textsf{\textit{Interaction}}, including aspects that arise in the ordinary operation of a user familiar with the system. These aspects assume expected behavior, not accounting for deficiencies such as errors or failures. The \textsf{\textit{Interaction}} category was found to encompass the following aspects:

\begin{itemize}
\item \textit{\underline{Algorithm.}} Users struggle to comprehend why a system generates a particular output, wanting to know the factors that influenced the computation. The output is unique to each user, therefore the programmed logic that is the same for all users is not included in this aspect (see aspect \textit{Design Decisions}). Example [sic]: ``In the last 3 months my credit went up a total of 10 points and then dropped down 7 points December 2. This doesn't make sense.''

\item \textit{\underline{Design Decision.}} Users wonder why things are a particular way (status quo) or not a certain way (counterfactual). It is not an output of the system that might be individual to each user, but the programmed logic, which the developers have agreed on. Hence, in contrast to the \textit{Algorithm} aspect, the \textit{Design Decisions} are the same for multiple (if not all) users. Example [sic]: ``why does the app force portrait mode?''

\item \textit{\underline{Signification.}} Users seek clarification on definitions, visual elements (such as symbols, colors, and highlighting), information visualizations, or related issues in order to understand the system's intended meaning. Example [sic]: ``I like this app, but when there may be something in red I just don't understand. Does it means something is wrong?''
\end{itemize}

The last category in the primary concerns is \textsf{\textit{Business}} category. It represents general \emph{Explanation Needs} that are not necessarily provoked during the interaction with a system. Further, aspects to be explained may be shaped by overarching business goals or specific project or process requirements~\cite{glinz_non-functional_2007}. Here we determined the following aspects:

\begin{itemize}

\item \textit{\underline{Mission.}} Users seek clarification on the system's purpose, utility, and vision, with a particular focus on specific features and the system as a whole. Example [sic]: ``Why do we need to access this app to get the information we used to get by phone from the doctor?''

\item \textit{\underline{Purchase \& Subscription.}} Users inquire about purchase or subscription matters, such as feature exclusivity in premium. This aspect only applies when there are multiple product lines with varying purchase or subscription plans. Example [sic]: ``Do I have to pay for it on all devices?''

\item \textit{\underline{Privacy.}} Users express privacy concerns regarding data collection, processing, and forwarding practices, as well as legal privacy rights and app permissions (e.g., GPS activation). If the inquiry is not focused on privacy but rather on the aspects that affect software decisions, it falls under the \textit{Algorithm} aspect. Example [sic]: ``Not sure why you need date of birth to register a navigation app, very suspicious as far as I'm concerned.''
\end{itemize}

Moving to the secondary concerns, we have \textsf{\textit{Dissatisfaction}} and \textsf{\textit{Errata}} categories. Accordingly, here the \emph{Explanation Need} is only the secondary concern of users, and there is a substantial underlying problem (at least in the user's perception) that is their primary concern. Overall, the \textsf{\textit{Dissatisfaction}} aspects are somewhat reproachful and the primary concern typically is a subjective deficiency from the user's point of view.

\begin{itemize}

\item \textit{\underline{Change.}} Users seek explanations for changes to a system, including modifications to the user interface or workflow. This aspect is more critical than genuine. However, it does not necessarily involve the need for re-learning the system, which is covered by the \textit{Instruction} aspect. Example [sic]: ``It just keeps getting worse. Why do you do this?''

\item \textit{\underline{Feature Gap.}} Users want to know why a feature is incomplete or missing. This aspect doesn't cover cases where a feature is not supported for an individual user's use case (see aspect \textit{Compatibility}). Example [sic]: ``Why would you have a database where you can only add and not edit or delete?''

\item \textit{\underline{Compatibility.}} Users are confused by a feature(s) not being supported or compatible with their use case. So, they are prevented from using a set of features due to external conditions that are not part of the system. This aspect excludes errors or failures. Example [sic]: ``Only big downfall is that USA account holders for some reason ... cannot use the boost feature. No clue why and no one has given answers to why it doesn't work.''
\end{itemize}

Finally, the \textsf{\textit{Errata}} category describes a situation with an undeniable objective deficiency such as an error or failure~\cite{avizienis_basic_2004} in the system. It differs with \textsf{\textit{Dissatisfaction}}, where the primary concern is a subjective deficiency in a user's eyes. We found the following aspects to be typical for \textsf{\textit{Errata}}:

\begin{itemize}

\item \textit{\underline{Fix.}} Users ask about fixes or workarounds to solve errors/ failures or ask whether errors/failures are known to the developers. Example [sic]: ``Anyone experiencing the same or know what to do about it?''.

\item \textit{\underline{Cause.}} Users ask for the underlying faults that cause errors, failures, or obviously erroneous outputs. They are interested in knowing the cause of the errors/failures to potentially attempt to fix them themselves. On the contrary, they do not ask for any support (see aspect \textit{Fix}). Example [sic]: ``Is it a loading problem or a glitch??''.

\item \textit{\underline{Confusing Message.}} Users feel misled by rare messages (such as uninformative or incongruous alerts) and assess the messages as incomplete, inaccurate, or erroneous. The messages can potentially be faulty explanations. Example [sic]: ``I constantly get warnings that I don't have enough shares to sell and I cannot find any solutions''.
\end{itemize}

\subsection{Discussion of Results}

\begin{table*}
\footnotesize
\centering
\caption{Overview of all datasets and annotations. \ens is used for the taxonomy creation (see \autoref{sec:taxonomy}). \pbr and \scrape are used to train and evaluate our explanation need detection approaches (see \autoref{sec:ml}).}
\label{tab:overviewApps}
        \vspace{-.1cm}
\renewcommand{\arraystretch}{0.9}
\begin{tabularx}{\linewidth}{c c r l l l r r c*{4}{>{\centering\arraybackslash}X}@{}}
\toprule
& & & & & & & & \multicolumn{5}{c}{Distribution of Expl. Needs} \\
\cmidrule(lr){9-13}
& & & Apps & Description & Source & Size & Expl. Needs & Tra & Int & Bus & Dis & Err\\ \midrule
        \cellcolor{blue!12} & \cellcolor{red!12} & 1. & Baby Tracker & Newborn Tracking & & 200 & 8 (4.0\%) & 3 & 1 & 1 & 3 & ~ \\
        \cellcolor{blue!12} & \cellcolor{red!12} & 2. & Experian Credit & Credit Reporting & ~ & 210 & 10 (4.8\%) & ~ & 5 & 1 & 3 & 1\\ 
        \cellcolor{blue!12} & \cellcolor{red!12} & 3. & FollowMyHealth & Patient Engagement & ~ & 229 & 11 (4.8\%) & 4 & ~ & 1 & 3 & 3 \\ 
        \cellcolor{blue!12} & \cellcolor{red!12} & 4. & Stock Master & Financial Trading & \cite{brunotte2022app} & 226 & 9 (4.0\%) & 3 & 2 & 1 & 3 & ~ \\ 
        \cellcolor{blue!12} & \cellcolor{red!12} & 5. & Here We Go & Navigation Software & ~ & 220 & 9 (4.1\%) & 1 & 1 & 1 & 3 & 3  \\ 
        \cellcolor{blue!12} & \cellcolor{red!12} & 6. & MiBand & Smartwatch Companion & ~ & 217 & 9 (4.1\%) & 2 & 2 & 1 & ~ & 4 \\ 
        \cellcolor{blue!12} & \cellcolor{red!12} & 7. & Waze & Navigation Software & ~ & 221 & 8 (3.6\%) & 1 & ~ & ~ & 4 & 3 \\ 
        \multirow{-8}{*}{\cellcolor{blue!12} \rotatebox[origin=c]{90}{\parbox{2cm}{\centering Tax-DS \\ \tiny (1730 reviews)}}} & \cellcolor{red!12} & 8. & Yazio & Nutrition Tracking & ~ & 207 & 12 (5.8\%) & 3 & 2 & ~ & 7 & ~ \\ \cline{3-13}
        & \cellcolor{red!12} & 9. & Unkown Apps & & \cite{maalej_automatic_2016} & 2449 & 108 (4.4\%) & 19 & 19 & 13 & 41 & 16 \\ \cline{3-13}
        & \cellcolor{red!12} & 10. & Amazon Prime & Video Streaming & & 100 & 10 (10.0\%) & ~ & 2 & 3 & 3 & 2 \\ 
        & \cellcolor{red!12} & 11. & AutoSleep & Sleep Tracking & ~ & 100 & 2 (2.0\%) & 1 & 1 & ~ & ~ & ~ \\ 
        & \cellcolor{red!12} & 12. & Disney+ & Video Streaming & ~ & 100 & 9 (9.0\%) & 1 & 1 & 3 & 2 & 2 \\ 
        & \cellcolor{red!12} & 13. & HotSchedules & Work Scheduling & ~ & 100 & 12 (12.0\%) & 3 & 1 & 5 & ~ & 3 \\ 
        & \cellcolor{red!12} & 14. & McDonald’s & Fastfood Companion & ~ & 100 & 15 (15.0\%) & ~ & 6 & 4 & 4 & 1 \\ 
        & \cellcolor{red!12} & 15. & Procreate Pocket & Digital Painting & ~ & 100 & 8 (8.0\%) & 5 & ~ & ~ & 3 & ~ \\ 
        & \cellcolor{red!12} & 16. & SkyView & Interactive Education & Ours & 100 & 9 (9.0\%) & 3 & 2 & 1 & 3 & ~ \\ 
        & \cellcolor{red!12} & 17. & Workoutdoords & Fitness Companion & ~ & 100 & 0 (0.0\%) & ~ & ~ & ~ & ~ & ~  \\ 
        & \multirow{-18}{*}{\cellcolor{red!12} \rotatebox[origin=c]{90}{\parbox{2cm}{\centering CrossVal-DS \\ \tiny (5078 reviews)}}} & 18. & YouTube & Social Network & ~ & 99 & 12 (12.1\%) & ~ & 3 & 2 & 6 & 1 \\
        & \cellcolor{orange!12} & 19. & WeChat & Social Network & ~ & 125 & 18 (14.4\%) & 2 & 9 & ~ & 3 & 4 \\
        & \cellcolor{orange!12} & 20. & Memrise & Language Learning & ~ & 122 & 1 (0.8\%) & 1 & ~ & ~ & ~ & ~ \\ 
        & \cellcolor{orange!12} & 21. & Duolingo & Language Learning & ~ & 118 & 2 (1.7\%) & ~ & ~ & ~ & 1 & 1 \\ 
        & \multirow{-4}{*}{\cellcolor{orange!12} \rotatebox[origin=c]{90}{\parbox{1.3cm}{\centering  Gen.-DS \\ \tiny (486 reviews)}}} & 22. & GitHub & Version Control & ~ & 121 & 3 (2.5\%) & 1 & 2 & ~ & ~ & ~ \\ \midrule
        & & & \textbf{Total} & ~ & ~ & \textbf{5564} & \textbf{285} (5.1\%) & \scriptsize \textbf{53} (18.6\%) & \scriptsize \textbf{59} (20.7\%)  & \scriptsize \textbf{37} (13.0\%) & \scriptsize \textbf{92} (32.3\%) & \scriptsize \textbf{44} (15.4\%) \\ \bottomrule
        \multicolumn{8}{l}{\scriptsize \textbf{Tra:} {Training},  \hspace{1pt} \textbf{Int:} { Interaction},  \hspace{1pt}\textbf{Bus:} { Business},   \hspace{1pt}\textbf{Dis:} { Dissatisfaction},   \hspace{1pt}\textbf{Err:} { Errata}}
\end{tabularx}
        \vspace{-.4cm}
\end{table*}

Through a rigorous study of app reviews, we have developed the \emph{Explanation Needs} taxonomy, which addresses \textit{RQ1} and provides a valuable resource for researchers and developers seeking to understand the concerns and requirements of end-users. By categorizing user needs in the taxonomy, we can better recognize and address various requirements in a more systematic manner, ultimately improving the quality, transparency, and user-friendliness of the application. The proposed taxonomy serves as an enabler, allowing for a more effective approach to addressing user needs and fostering a deeper understanding of the end-user experience. As such, the \emph{Explanation Needs} taxonomy has significant implications for app development and can contribute to the development of more explainable systems that better meet the needs of users.

With the \emph{Explanation Needs} taxonomy, we were able to tackle the \textit{RQ2}, which aimed to gain a more statistical view of the types of \emph{Explanation Needs} expressed in app reviews. So we applied the taxonomy to multiple sets of data, composed of 5,564 reviews in total. Table~\ref{tab:overviewApps} provides an overview of all the datasets used in this paper. As discussed in Section~\ref{sec:taxonomy}, the taxonomy extraction was based on the \ens\ and the final labeling was achieved through several rounds of cross-checking to ensure the validity and reliability of our findings. 

However, to gain deeper insights into the types of information and \emph{Explanation Needs} in the app reviews and to further assess the coverage and applicability of our taxonomy, we also labeled the reviews of our extended datasets, which we create for classifier implementation and validations (see Section~\ref{subsec:corpus} for more details). The labeling process of the rest of the data (i.e., the app reviews 9 to 22 in Table~\ref{tab:overviewApps}) was carried out after consolidating the taxonomy and codebook, the latter of which provides complete information on inclusion and exclusion criteria, as well as typical and atypical examples. Following this, a single coder categorized the app reviews in \pbr and \scrape that had already been labeled as \emph{Explanation Needs} (see Section~\ref{sec:ml} for more details).

Besides the description of the apps, source and number of reviews, Table~\ref{tab:overviewApps} provides a breakdown of the distribution of different types of \emph{Explanation Needs} per app. It shows the number of occurrences of each type of \emph{Explanation Needs} for each app, as well as the total number and percentage of \emph{Explanation Needs} across all apps. By examining this table, we can answer the \textit{RQ2} by identifying the areas where users require the most explanations. This analysis can help shed light on the nature and extent of \emph{Explanation Needs} in app reviews.

For example, it shows that the majority of cases fall under the \textit{Primary Concerns} category, accounting for 52.3\% of all app reviews. This implies that users' primary issue with the app is their lack of understanding and knowledge, without any substantial problems aside from it. This finding highlights the importance of addressing users' primary concerns and providing sufficient explanations to enhance their overall understanding of the app's functionality. Furthermore, the \textsf{\textit{Interaction}} category is the most frequent type within the \textit{Primary Concerns} and accounts for 20.7\% of the total number of \emph{Explanation Needs} across all apps. This means that a significant proportion of user feedback in app reviews is related to ordinary interaction with the system. As users engage with the app, they may encounter unexpected behaviours, have questions about design decisions, or need clarification on the meaning of certain visual elements or notions. Accordingly, it is not surprising to have a relatively high number of \textsf{\textit{Interaction}} types since these issues could arise regardless of the app's specific functionality, and, therefore, could be relevant to a wide range of users. Additionally, the \textsf{\textit{Interaction}} category may be particularly salient to users, as it directly affects their experience using the app, and they may be more likely to leave reviews on these types of issues. Similarly, the \textsf{\textit{Training}} category stands out with the second-highest percentage of \emph{Explanation Needs} in the primary concern, accounting for 18.6\% of all \emph{Explanation Needs}, indicates that users frequently encounter difficulties in understanding how to use certain features or functionalities of the app. This finding highlights the importance of providing concise instructions or tutorials to help users learn how to use the app effectively.
Overall, the high percentage of \textsf{\textit{Training}} and \textsf{\textit{Interaction}} indicates that the app's user interface or design could be improved. Our results hence may suggest that the application design and development should primarily focus on the usability of the apps by making them more intuitive and user-friendly. 

Another interesting observation is that the \textsf{\textit{Dissatisfaction}} category, which is classified as a secondary concern, has the highest percentage of \emph{Explanation Needs} at 32.3\%. This could be attributed to its subjective nature, as the primary concern of this category is a perceived deficiency from the user's point of view, which may be difficult to address directly. Additionally, this deficiency is not necessarily related to a specific bug or technical issue, but rather a mismatch between the user's expectations and the app's performance or features. This finding suggests that users are more likely to express their discontentment and frustration in reviews. Last but not least, our qualitative analysis also reveals an important insight. We found that although app reviews provide a wealth of information about users' \emph{Explanation Needs}, the proportion of reviews that contain such information is relatively low, at only 5.1\%. This indicates a need for more efficient and automated techniques to extract useful content from reviews. Therefore, our study has motivated us to pursue our second contribution, which is described in more detail in Section~\ref{sec:ml}. By developing machine learning-based approaches to extract \emph{Explanation Needs} from reviews, we hope to improve the efficiency and effectiveness of analyzing large volumes of user feedback.

\subsection{Threats to Validity}


A potential threat to internal validity is the use of quantitative coding, which can be interpretive and subjective. This means that our analysis may be influenced by our own biases or assumptions, which could affect the accuracy of our findings. Poor English and typos in some reviews can also lead to inaccurate conclusions, but we made a conscious effort to evaluate unintelligible reviews. In addition, a threat to external validity could be survivorship bias, as our results may not be representative of those with low technological literacy, as they may be less likely to write and publish app reviews in the first place. Also, the \ens we used in our taxonomy extraction is relatively small, with only a few cases of \emph{Explanation Needs} observed (4.6\% as shown in Table~\ref{tab:overviewApps}). Accordingly, it might limit the generalizability of our taxonomy categories. However, to mitigate the potential threat of a small sample, we conducted a thorough and saturated coding process and verified the validity of our taxonomy categories on an extended dataset.


\section{Automatic Detection of Explanation Needs}\label{sec:ml}

\subsection{Corpora Creation}
\label{subsec:corpus}
To determine the best method for detecting \emph{Explanation Needs} in a structured way, we follow the recommendations by Dell’Anna~et~al.~\cite{DellAnna2022}. They stress that the results of a simple cross-validated experiment do not allow to draw definite conclusions about the performance of a classifier in an operational context. In other words, we cannot necessarily infer from such an experiment whether the classifier is able to generalize and is thereby suitable for use on unseen data in practice. Hence, we evaluate our approaches on two datasets:

\noindent \textbf{CrossVal-DS.} We use this dataset to train and compare all models applying 10-fold cross-validation. The main purpose of \pbr is to compare the performance of different \ac{NLP} classifiers and to select the best-performing method. It includes all reviews of \ens created in Section~\ref{sec:taxonomy}. However, this dataset with 77 \emph{Explanation Needs} is not sufficient for training an \ac{NLP} classifier. Accordingly, we extend the dataset with further reviews and manually label them with respect to the tags ``explanation need'' and ``no explanation need''. We make use of a dataset collected by Maalej et al.~\cite{maalej_automatic_2016} that has already been utilized in the \ac{RE} community to classify app reviews into problem reports, inquiries, and irrelevant ones~\cite{henao_transfer_2021, stanik_classifying_2019}. Additionally, we collect further app reviews from 9 popular apps, using custom Python web scraping tools for the Apple App Store\footnote{\url{https://pypi.org/project/app-store-scraper/}} and Google Play Store\footnote{\url{https://pypi.org/project/google-play-scraper/}}. For each of the apps, we scraped as many reviews as possible and then drew a random sample of 100 reviews to include an equal-sized subset of the reviews per app. A detailed overview of \pbr is provided in Table~\ref{tab:overviewApps}. In total, \pbr comprises 5,078 reviews of which 261 contain \emph{Explanation Needs} (5.14\%).\looseness=-1

\noindent \textbf{General-DS.} To investigate the generalizability of the best-performing classifier, we apply it to a set of unseen reviews that are not associated with any of the apps contained in \pbr. Specifically, we scrape and annotate reviews about the four randomly selected apps called WeChat, Memrise, Duolingo, and GitHub (see Table~\ref{tab:overviewApps}). The main purpose of \scrape is to report the performance of our best classifier in a realistic setting. In total, \scrape comprises 486 reviews of which 24 contain \emph{Explanation Needs} (4.94\%).

\begin{table}
    \centering
    \caption{Inter-annotator agreement statistics ($n=485$).}
    \label{tab:interRaterDS4}
    \begin{tabular}{@{}rrr@{}}
        \toprule
           & \multicolumn{2}{c}{Rater 1} \\ 
         Rater 2 & no expl.\ need & expl.\ need \\
         no expl.\ need & 448 & 17 \\
         expl.\ need & 7 & 13 \\ \midrule
         Agreement  & \multicolumn{2}{c}{95.05\%} \\
         Cohen's Kappa  & \multicolumn{2}{c}{0.495} \\
         Gwet's AC1  & \multicolumn{2}{c}{0.945} \\
        \bottomrule
    \end{tabular}
\end{table}

\subsection{Annotation Validity}
To verify the reliability of our annotations, we calculated the inter-annotator agreement in terms of Cohen's Kappa~\cite{cohen60}. We involved a total of four annotators in the creation of \pbr and \scrape and assessed the inter-rater reliability on the basis of 485 reviews that each have been labeled by two out of the four annotators. 
In case of a high imbalance of ratings, Cohen's Kappa is low and indicates poor inter-rater reliability even if there is a high agreement between the raters (Kappa paradox~\cite{FEINSTEIN1990}). Thus, Cohen's Kappa is not meaningful in such scenarios. Consequently, Cohen's Kappa should always be reported together with the percentage of agreement and other paradox-resistant measures (e.g., Gwet's AC1 measure~\cite{gwet}). We calculated all measures (see Table~\ref{tab:interRaterDS4}) using the cloud-based version of AgreeStat\footnote{\url{https://www.agreestat.com/}}. Cohen's Kappa and Gwet's AC1 can both be interpreted using the taxonomy developed by Landis and Koch~\cite{landis77}: values $\leq$ 0 as indicating no agreement and 0.01–0.20 as none to slight, 0.21–0.40 as fair, 0.41–0.60 as moderate, 0.61–0.80 as substantial, and 0.81–1.00 as almost perfect agreement. Table~\ref{tab:interRaterDS4} demonstrates that the inter-rater agreement of our annotation process is reliable as we achieve an average percentage of agreement of 95\%. Despite a high agreement of over 90\%, Cohen's Kappa yields a relatively low value, which \textit{paradoxically} suggests only moderate agreement. A more meaningful assessment is provided by Gwet's AC1 as it did not fail in the case of prevalence and remains close to the percentage of agreement. The achieved Gwet's AC1 of 0.945 indicates a nearly perfect agreement. Therefore, we assess \pbr and \scrape as reliable and suitable for the implementation and evaluation of our \emph{Explanation Need} detection approach.

\subsection{Methods}
We define the detection of \emph{Explanation Needs} as a binary classification problem, in which we are given a certain review $\mathcal{X}$ and we are required to produce a nominal label $y \in \mathcal{Y} = \{\text{explanation need}, \text{no explanation need}\}$. Since app store reviews are written in natural language, we build our classifier based on different methods established for \ac{NLP}.

\noindent\textbf{Rule-based Approach.}
Instead of using a random classifier as the baseline approach, we involve simple regex expressions for the detection of \emph{Explanation Needs}. We iterate through all reviews in the test set and check if a question mark or the word ``why'' is contained. We hypothesize that both expressions might be a feasible indicator for the presence of an \emph{Explanation Need}. Following this assumption, we classify a review as an \emph{Explanation Need} if it contains at least one of the two expressions and vice versa. 

\noindent\textbf{Machine Learning-based Approach.}
We investigate the use of \textit{supervised} \ac{ML} models that learn to predict \emph{Explanation Needs} based on a labeled dataset. Specifically, we employ established binary classification algorithms: \ac{NB}, \ac{SVM}, \ac{RF}, \ac{DT}, \ac{LR}, \ac{AB}, and \ac{KNN}. To determine the best hyperparameters for each binary classifier, we apply Grid Search, which fits the model on every possible combination of hyperparameters and selects the most performant. We use two different methods as word embeddings: \ac{BoW} and \ac{TF-IDF}. In Table~\ref{tab:classificationResults} we report the classification results of each algorithm as well as the best combination of hyperparameters.

\noindent\textbf{Deep Learning-based Approach.}
With the rise of \ac{DL}, more and more researchers are using \ac{DL} models for \ac{NLP} tasks. In this context, the \ac{BERT} model~\cite{devlin_bert_2019} is prominent and has already been used for question answering and named entity recognition. \ac{BERT} is pre-trained on large corpora and can therefore easily be fine-tuned for any downstream task without the need for much training data (Transfer Learning). In our paper, we make use of the fine-tuning mechanism of \ac{BERT} and investigate to which extent it can be used for the detection of \emph{Explanation Needs}. First, we tokenize each app store review. \ac{BERT} requires input sequences with a fixed length (maximum 512 tokens). Therefore, for reviews that are shorter than this fixed length, \ac{PAD} are inserted to adjust all reviews to the same length. Other tokens, such as the \ac{CLS}, are also inserted in order to provide further information on the review to the model. CLS is the first token in the sequence and represents the whole review (i.e., it is the pooled output of all tokens of a review). For our classification task, we mainly use this token because it stores the information of the whole review. We feed the pooled information into a single-layer feedforward neural network that uses a softmax layer, which calculates the probability that a review contains an \emph{Explanation Need} or not.

\subsection{Evaluation Procedure}
\pbr is strongly imbalanced as only 261 are positive samples. To avoid the class imbalance problem, we apply Random Under Sampling. We randomly select reviews from the majority class and exclude them from the dataset until a balanced distribution is achieved. Our final dataset consists of 522 reviews of which 261 contain an \emph{Explanation Need} and the other 261 do not. We follow the idea of cross-validation and divide the dataset into a training, validation, and test set. We opt for 10-fold cross-validation as a number of studies have shown that a model that has been trained this way demonstrates low bias and variance~\cite{James13}. Please note that undersampling stands in conflict with our goal to understand how well our classifier generalizes and performs in a realistic setting. Hence, we do not undersample \scrape allowing us to report our final results on a realistically distributed test corpus.

We use standard metrics for evaluating our approaches, such as Precision, Recall, and a weighted F-measure. Since a single run of a \textit{k}-fold cross-validation may result in a noisy estimate of model performance, we repeat the cross-validation procedure five times and average the scores from all repetitions. 
Since our classifier is supposed to assist development teams by detecting relevant \emph{Explanation Needs} in reviews automatically, we favor Recall over Precision. A high Recall corresponds to a greater degree of automation of \emph{Explanation Need} detection because it is easier for users to discard \ac{FP} than to manually detect \ac{FN}. Consequently, we seek high Recall to minimize the risk of missed \emph{Explanation Needs} and acceptable Precision to ensure that the development teams are not overwhelmed by \ac{FP}. To attain a accumulated, single metric from Precision and Recall, the simple F-Measure (F1) is frequently used in binary classification tasks. It is defined as the harmonic mean between Precision and Recall, and thus assigns equal importance to both metrics. To account for our preference for Recall over Precision, it is imperative to make adjustments to the way in which the two metrics are weighted. We evaluate our approaches based on a weighted F-Measure:
\begin{myequation}
    F_{\beta} = (1+\beta^2) \cdot \frac{Precision \cdot Recall}{(\beta^2 \cdot Precision) + Recall}
\end{myequation}

\noindent where $\beta$ is the ratio to which Recall is more important than Precision~\cite{hayes_advancing_2006}. Berry~\cite{berry_empirical_2021} defines $\beta$ as follows:
\begin{myequation}
    \beta = \frac{time_a \cdot \lambda}{time_v}
\end{myequation}

\noindent where $time_a$ is the average time that a human would need to assess an artifact manually (i.e., the time spent by a human determining whether a particular review is an \emph{Explanation Need} or not), and $time_v$ is the average time that a human would need to verify whether a positive detection by a tool is actually a True Positive (i.e., the time spent by a human neglecting a \ac{FP} detection of an \emph{Explanation Need}). Further, $\lambda$ is the inverse of the share of relevant artifacts within all artifacts. In other words, $\lambda$ is the average number of artifacts that an analyzer would need to investigate in order to find a single relevant artifact. In our case, $\lambda$ is calculated as follows:
\begin{myequation}
    \lambda = (\frac{285}{5564})^{-1} \approx 19.52
\end{myequation}

\noindent because we identified a total of 285 \emph{Explanation Needs} in our dataset of 5,564 reviews. Thus on average, one out of 19.52 app reviews contains an \emph{Explanation Need}. Since the time required to vet a single answer of our classifier is no more than the time required to manually check if an app review contains an \emph{Explanation Need}, the weight ratio $\beta$ is equal to $\lambda$. Hence, we define $\beta$ as 19.52.

\begin{table*}
    \centering
    \caption{Experimental Classification Results ($\beta$ = 19.52)}
            \vspace{-.1cm}
    \label{tab:classificationResults}
\renewcommand{\arraystretch}{0.9}
    \begin{tabularx}{1 \linewidth}{@{} l l X r r r r r r r@{}}
        \toprule
        & & & \multicolumn{3}{c}{\textbf{Expl. Need}} & \multicolumn{3}{c}{\textbf{Not Expl. Need}} & \\
        & & & \multicolumn{3}{c}{(Support: 53)} & \multicolumn{3}{c} {(Support: 53)} & \\ \cmidrule(lr){4-6} \cmidrule(lr){7-9}
        Method & & Best hyperparameters & Rec & Pre & {$F_\beta$} & Rec & Pre & {$F_\beta$} & Mac-{$F_\beta$}\\ \midrule
        \textbf{Rule-based} & ~ & include: '?' OR ‚why‘ & 0.92 & 0.94 & 0.92 & 0.94 & 0.92 & 0.94 & 0.93 \\  \midrule
        
        & \ac{NB} & alpha: 1, fit\_prior: False, embed: TF-IDF & 0.81 & 0.63 & 0.81 & 0.52 & 0.73 & 0.52 & 0.66 \\
        
        ~ & \ac{SVM} & C: 1, gamma: 0.001, kernel: linear, embed: TF-IDF & 0.74 & 0.72 & 0.74 & 0.71 & 0.73 & 0.71 & 0.73 \\ 
        ~ & \ac{RF} & criterion: entropy, max\_features: auto, n\_estimators: 500, embed: TF-IDF & 0.72 & 0.78 & 0.72 & 0.79 & 0.74 & 0.79 & 0.75 \\ 
        ~ & \ac{DT} & criterion: gini, max\_features: log2, splitter: best, embed: TF-IDF & 0.60 & 0.59 & 0.60 & 0.57 & 0.59 & 0.57 & 0.58 \\ 
        ~ & \ac{LR} & C: 1, solver: newton-cg, embed: TF-IDF & 0.74 & 0.71 & 0.74 & 0.70 & 0.73 & 0.70 & 0.72 \\ 
        ~ & \ac{AB} & algorithm: SAMME, n\_estimators: 50, embed: TF-IDF & 0.67 & 0.78 & 0.67 & 0.81 & 0.71 & 0.81 & 0.74 \\ 
        ~ & \ac{KNN} & algorithm: ball\_tree, n\_neighbors: 20, weights: uniform, embed: TF-IDF & 0.76 & 0.61 & 0.76 & 0.52 & 0.69 & 0.52 & 0.64 \\ 
        \textbf{\ac{ML}-based} & \ac{NB} & alpha: 1, fit\_prior: True, embed: BoW & 0.73 & 0.63 & 0.73 & 0.58 & 0.68 & 0.58 & 0.65 \\ 
        ~ & \ac{SVM} & C: 100, gamma: auto, kernel: rbf, embed: BoW & 0.70 & 0.75 & 0.70 & 0.77 & 0.72 & 0.77 & 0.74 \\ 
        ~ & \ac{RF} & criterion: entropy, max\_features: auto, n\_estimators: 500, embed: BoW & 0.72 & 0.78 & 0.72 & 0.79 & 0.74 & 0.79 & 0.76 \\ 
        ~ & \ac{DT} & criterion: gini, max\_features: log2, splitter: best, embed: BoW & 0.61 & 0.64 & 0.61 & 0.65 & 0.62 & 0.65 & 0.63 \\ 
        ~ & \ac{LR} & C: 1, solver: liblinear, embed: BoW & 0.70 & 0.75 & 0.70 & 0.77 & 0.72 & 0.77 & 0.73 \\ 
        ~ & \ac{AB} & algorithm: SAMME, n\_estimators: 200, embed: BoW & 0.69 & 0.78 & 0.69 & 0.81 & 0.73 & 0.81 & 0.75 \\ 
        ~ & \ac{KNN} & algorithm: ball\_tree, n\_neighbors: 16, weights: distance, embed: BoW & 0.38 & 0.68 & 0.38 & 0.82 & 0.57 & 0.82 & 0.60 \\  \midrule
        \textbf{\ac{DL}-based} & BERT & batch\_size: 16, learning\_rate: 2e-05, weight\_decay: 0.01 & 0.94 & 0.93 & 0.94 & 0.93 & 0.94 & 0.93 & 0.93 \\
        \bottomrule
    \end{tabularx}
            \vspace{-.4cm}
\end{table*}

\subsection{Experimental Results}
In the following, we describe the results of our experiments. First, we compare the performance of different \ac{NLP} classifiers on \pbr. Second, we investigate the generalizability of the best-performing method on \scrape.

\paragraph{Selection of Best-Performing Method}
Table~\ref{tab:classificationResults} reveals that our shallow rule-based approach shows a strong performance in detecting \emph{Explanation Needs}. It achieves a high $F_{19.52}$ score for both classes and is able to demarcate between reviews that contain \emph{Explanation Needs} and those that do not. In comparison, all \ac{ML}-based approaches exhibit a significantly poorer performance. For example, \ac{DT} trained on \ac{TF-IDF} embeddings achieves a Macro-$F_{19.52}$ score of 58\% (deterioration of 35\% compared to the baseline approach). The best performance in this category is achieved by \ac{RF} trained on \ac{BoW} embeddings with a Macro-$F_{19.52}$ score of 76\%. Our experiment shows that the choice of sentence embedding has no significant effect on the performance of the \ac{ML}-based approaches. Most of the approaches achieve a Macro-$F_{19.52}$ score of about 70\% regardless of the applied sentence embedding. Our fine-tuned \ac{BERT} model, on the other hand, shows a considerably stronger performance and achieves a Macro-$F_{19.52}$ score of 93\%. Interestingly, despite its rich language understanding, the \ac{BERT} model fails to outperform our simple rule-based approach. In fact, both approaches achieve the same Macro-$F_{19.52}$ score and posses consequently the same predictive power. Our experiments thus show that both approaches are suitable for identifying \emph{Explanation Needs} in \pbr. To investigate the generalizability of the rule-based approach and the \ac{BERT} model, we apply both approaches to a larger set of unseen reviews written for other apps contained in \scrape.\looseness=-1

\paragraph{Generalizibility of Best-Performing Method}
When applied to unseen data, both approaches show a clear performance drop in the detection of \emph{Explanation Needs} (see Table~\ref{tab:generalizabilityClassificationResults}). While both approaches continue to show very high $F_{19.52}$ scores for the ``no explanation need'' class, the $F_{19.52}$ score for the ``explanation need'' class has decreased significantly. The largest performance drop is evident in the rule-based approach, which only shows an $F_{19.52}$ score of 67\% in detecting explanation needs across all reviews of all four apps. Similarly, the trained \ac{BERT} model fails to match the very good $F_{19.52}$ score of 94\% that it could achieve when applied to the balanced training set. Instead, it achieves a score of 79\% on the unseen data, which corresponds to a decrease of 15\%. Overall, the \ac{BERT} model outperformed the rule-based approach and achieved a significantly better Macro-$F_{19.52}$ score of 86\%.

The higher Macro-$F_{19.52}$ score is mainly attributable to the fact that the \ac{BERT} model shows a significantly better Recall with regard to the \emph{Explanation Need} class. In other words, the \ac{BERT} model identified more \emph{Explanation Needs} in the reviews than the rule-based system. Our experiment demonstrates that this performance deviation does not depend on a specific app about which the respective reviews were written. In fact, when applied to the reviews about WeChat, Duolingo and Github, the \ac{BERT} model exhibits better performance. In the case of the reviews about Memrise, it achieves the same Recall as the rule-based approach. Both the rule-based approach and the \ac{BERT} model show the most significant performance loss with regard to Precision and generate a great number of \ac{FP}. Using both approaches, two of three reviews that are supposed to contain an \emph{Explanation Need} are \ac{FP}s, causing high filtering costs for practitioners.

\begin{table}
    \centering
    \caption{Experimental classification results on unseen data (RB = rule-based baseline, DL = deep-learning-based, $\beta$ = 19.52)}
     \vspace{-.1cm}
\label{tab:generalizabilityClassificationResults}
\renewcommand{\arraystretch}{0.9}
    \begin{tabularx}{\linewidth}{@{}llrrrrrrr@{}}
        \toprule
        & & \multicolumn{3}{c}{\textbf{Expl. Need}} & \multicolumn{3}{c}{\textbf{No Expl. Need}} & \\ \cmidrule(lr){3-5} \cmidrule(lr){6-8}
        App &  & Rec & Pre & $F_\beta$ & Rec & Pre & $F_\beta$  & Mac-$F_\beta$ \\ \midrule
        \textbf{WeChat} & RB & 0.67 & 0.41 & 0.67 & 0.84 & 0.94 & 0.84 & 0.75 \\ 
        ~ & DL & 0.72 & 0.41 & 0.72 & 0.82 & 0.95 & 0.82 & 0.77 \\ 
        \textbf{Memrise} & RB & 1.00 & 0.33 & 0.99 & 0.98 & 1.00 & 0.98 & 0.99 \\ 
        ~ & DL & 1.00 & 0.25 & 0.99 & 0.98 & 1.00 & 0.98 & 0.98 \\ 
        \textbf{Duolingo} & RB & 0.50 & 0.33 & 0.50 & 0.98 & 0.99 & 0.98 & 0.74 \\ 
        ~ & DL & 1.00 & 0.40 & 1.00 & 0.97 & 1.00 & 0.97 & 0.99 \\ 
        \textbf{GitHub} & RB & 0.67 & 0.33 & 0.66 & 0.97 & 0.99 & 0.97 & 0.82 \\ 
        ~ & DL & 1.00 & 0.30 & 0.99 & 0.94 & 1.00 & 0.94 & 0.97 \\ \midrule 
        \textbf{Total} & RB & 0.67 & 0.39 & 0.67 & 0.95 & 0.98 & 0.95 & 0.81 \\ 
        ~ & DL & 0.79 & 0.37 & 0.79 & 0.93 & 0.99 & 0.93 & 0.86 \\ 
        \bottomrule
    \end{tabularx}
\end{table}

\subsection{Discussion of Results}
Our experiments show that the rule-based approach achieves the same performance as the \ac{BERT} model when evaluated on \pbr, but performs worse when applied to unseen data. The rule-based approach fails to recognize more than 30\% of the \emph{Explanation Needs} and seems to generalize less effective than the \ac{BERT} approach. When analyzing the data in \scrape, we see that the detection of \emph{Explanation Needs} cannot be broken down to the presence of questions and question words. \emph{Explanation Needs} do not necessarily contain question marks or question words. In many cases, questions are formulated but question marks are not included: \enquote{Would you please keep us updated on what's going on. I have several texts and don't know how to keep them. Don't want to lose it.} The \ac{BERT} model understands the semantics of sentences better and dependents less on the sentence's syntax. The rule-based approach could be extended by adding more interrogatives (e.g., \textit{how}) and interrogative verbs (e.g., \textit{don't understand}) to enhance the Recall of the approach, however, this may lead to an unreasonable increase in \ac{FP}s. The resulting filtering effort would diminish the use of the approach in practice. 

From a critical point of view, our best classifier does not perform flawlessly. It does not identify all \emph{Explanation Needs} in \scrape and predicts a number of \ac{FP}s. We argue that the recall value needs to be improved above 90\% to qualify the approach for practical use. Otherwise, the practitioners would have to go through the reviews manually to detect false negatives, which is time-consuming given the high number of reviews and the fact that \emph{Explanation Needs} rarely occur. The achieved precision value of 37\% is not optimal, but in our view still justifiable. It is much easier for the practitioner to neglect two false positives from 3 reviews predicted as \emph{Explanation Needs} than to go through 20 reviews manually to discover a single \emph{Explanation Needs}. 

Our classifier marks a first step toward automatic \emph{Explanation Need} detection. Further studies should focus on optimizing the classifier in terms of recall. We hypothesize that the extension of the training set and the use of further language models might be beneficial. So far, we have only focused on the \ac{BERT}-Base model~\cite{devlin_bert_2019}, although other studies~\cite{FISCHBACH23} show that alternative models such as RoBERTa can achieve even better performance. To assist practitioners in filtering \ac{FP}s, it may also be useful to have the classifier mark the specific clause in each review that has caused the review to be categorised as \emph{Explanation Needs}~\cite{winkler17}. This will help practitioners to understand the inner workings of the classifier and also increase its acceptance. 

\subsection{Threats to Validity}
A threat to internal validity are the annotations themselves as an annotation task is subjective to a certain degree. To minimize the bias of the annotators, we performed two mitigation actions: First, we conducted a workshop prior to the annotation process to ensure a common understanding of \emph{Explanation Needs}. Second, we assessed the inter-rater agreement by using multiple metrics (Gwet's AC1 etc.). 
Despite our efforts to make the labeling process as transparent and systematic as possible, there may still be some variability in the resulting gold standard, e.g., misinterpretation of the users' intention, blurred boundaries between the categories, too broad or too narrow judgement, or human mistakes. Using the adjusted $F_\beta$-score as an evaluation metric poses a threat to construct validity. We used an adjusted $\beta$ value of 19.52, which was calculated based on the frequency of \textit{Explanation Need} occurrences in app reviews. This value is in the order of $\beta$ values calculated for other ``needle in the haystack'' tasks~\cite{berry_empirical_2021}. However, it is possible that the value may deviate when calculated based on another dataset. Our results have shown that generalization of our tested classifiers is fairly moderate when applied to unseen, dissimilar test data. This may indicate that more data is needed to train a classifier that generalizes better. 
Lastly, app reviews are not the only relevant source of user feedback~\cite{Nayebi2018}. 


\section{Conclusion}

This work is a further step towards user-centered explainability engineering.
It contributes to a better understanding of users' \emph{Explanation Needs} and lays the foundation for future research and development in this area. The proposed taxonomy of \emph{Explanation Needs} provides a rigorous approach for extracting explainability requirements from app reviews, ensuring that they meet users' expectations. In addition, our approach represents the first step towards automatic explanation need detection and reduces the manual effort required by engineers and researchers to identify \emph{Explanation Needs} in reviews. To facilitate practical use of the approach, it needs to be optimized for recall so that practitioners can efficiently focus on eliciting valid \emph{Explanation Needs}. Finally, our published set of manually labeled app reviews will enable researchers in the field to improve their own models and approaches for detecting \emph{Explanation Needs}. 

\section*{Acknowledgements} 
This work was funded by the Deutsche Forschungsgemeinschaft (DFG, German Research Foundation) under Grant No.: 470146331, project softXplain (2022-2025).


\bibliographystyle{IEEEtran}
\bibliography{references}

\end{document}